\documentclass[usenatbib,onecolumn,useAMS]{mn2e}
\usepackage{amssymb,epsfig,rotating}

\title[Jets in YSOs and AGN]
{A comparison of the acceleration mechanisms in YSO and AGN jets}
\author[Price, Pringle and King]{D.J. Price,$^1$ J.E. Pringle,$^1$ A.R. King,$^2$ \\
$^1$Institute of Astronomy, Madingley Rd, Cambridge, CB3 0HA, UK \\
$^2$Department of Physics and Astronomy, University of Leicester, Leicester, LE1
7RH, UK}
\date{Submitted: 8th August 2002}

\pagerange{\pageref{firstpage}--\pageref{lastpage}} \pubyear{2002}

\newcommand{\pder}[2]{\frac{\partial #1}{\partial #2}}

\begin{document}
\label{firstpage}
\maketitle

\begin{abstract}
We examine the hypothesis that there exists a simple scaling between
the observed velocities of jets found in Young Stellar Objects (YSOs)
and jets found in Active Galactic Nuclei (AGN). We employ a very
simplified physical model of the jet acceleration process. We use
time-dependent, spherically symmetric wind models in Newtonian and
relativistic gravitational fields to ask whether the energy input
rates required to produce the jet velocities observed in YSOs (of
about 2 $\times$ the escape velocity from the central object) can also
produce AGN jet velocities (Lorentz factors of $\gamma \sim$ 10). Such
a scaling would be expected if there is a common production mechanism
for such jets. We demonstrate that such a scaling does exist, provided
that the energy input process takes place sufficiently deep in the
gravitational potential well, enabling physical use to be made of the
speed of light as a limiting velocity, and provided that the energy
released in the accretion process is imparted to a small fraction of
the available accreting material.
\end{abstract}

\begin{keywords}
galaxies: jets -- ISM: jets and outflows -- stars: winds, outflows --
relativity
\end{keywords}

\section{Introduction}
Highly collimated jets are observed in a variety of astrophysical
objects. They have been found in quasars, active galactic nuclei
(AGN), stellar binaries, planetary nebulae, young stellar objects
(YSOs) and young pulsars (\citealt{f98}; \citealt{l99};
\citealt{rb01}; \citealt{g01}). However, despite a large theoretical
effort, there is still no agreement as to the mechanisms which give
rise to the acceleration and to the collimation of these jets
(\citealt{b93,b00,b02}; \citealt{p93}; \citealt{f98};
\citealt{l99}). In this paper we shall restrict our discussion mainly
to jets in AGN and in YSOs, although our findings will of course have
relevance to other areas.  In both AGN and YSOs, inflow of matter is
thought to be occurring through an accretion disc, and it is this
process which is thought to power the jets. Also in these objects, the
absence of substantial thermal emission implies that the jets are not
simple hydrodynamic flows, powered by thermal pressure (discussed for
example by \citealt{br74}; \citealt{k82}). For this reason it is
usually assumed that the acceleration mechanism is associated with
magnetic fields.  Because the first collimated jets to be discovered
were the relativistic radio jets from galactic nuclei, presumably
powered by accretion onto the central black hole \citep{r84}, all the
early models of jets involved the generation of jets from a central
black hole \citep{f98}. Indeed the frequently invoked Blandford-Znajek
mechanism \citep{bz77} for such jet formation requires the tapping of
rotational energy from a rapidly spinning black hole. It is clear,
however, that such exotic processes are of little relevance to the
generation of jets around young stellar objects. This leaves jet
theorists with a dilemma. Do we argue that the jet generation
mechanisms in AGN and YSOs are unrelated, and that all the various jet
formation models put forward so far are correct when applied to the
right object? Or can we apply Occam's razor, and argue that all jets
are produced by fundamentally the same mechanism, and that essentially
the same theory can be applied in both cases, when the appropriate
scalings are applied (\citealt{k86}; \citealt{p93};
\citealt{l97,l99})?

The jet acceleration process by a spinning disc is often envisaged as
being due to centrifugal acceleration of disc material by large scale
poloidal field lines threading the disc (\citealt{bp82};
\citealt{pn86}). This idea is exemplified by the magnetic wind
solution of \citet{bp82}, and has been demonstrated in a number of
numerical simulations (\citealt{op97,op99}; \citealt{ops97};
\citealt{kms98}; \citealt{kea00}). It appears to be a fairly generic
observed property of collimated jets that the jet velocities are
comparable to the escape velocities from the central gravitating
objects. In the YSO jets, the intrinsic jet velocity is hard to
measure, because in order to see the jet it needs to have interacted
with some of the surrounding material, and so to have been slowed to
some extent. Thus it is difficult to measure the core of the jet
directly. Typically the jet velocities at various part of the flow are
inferred by modelling the velocity structures in the neighbourhood of
HH objects (which are basically shocks within the jet) and from the
observed proper motions of the HH objects, which of course give lower
limits to the jet velocity \citep{rb01}. Such considerations indicate
that typical YSO jet velocities are in the range $v_{\rm jet} \sim
300-500$ km/s (\citealt{em98}; \citealt{mea98}; \citealt{bea01};
\citealt{hea01}; \citealt{rea02}; \citealt{bea02}; \citealt{pea02})
compared to the escape velocity from a typical young star (mass 1
M$_\odot$, radius 5 R$_\odot$; \citealt{tlb99}) of $v_{\rm esc} \sim
270$ km/s. The AGN jets are relativistic, appropriate for a velocity
of escape from close to a black hole, and appear observationally to
have relativistic gamma-factors around $\gamma_{\rm jet} \sim 5-10$
(\citealt{up95}; \citealt{bsm99}), although arguments for higher
values ($\gamma_{\rm jet} \sim 10-20$) have been made on theoretical
grounds \citep{gc01}. It is clear from this that scale-free models for
driving outflows from discs, such as the model of \citet{bp82}, are
lacking in the basic respect that the observed disc-launched jets
appear to know of the scale associated with, and thus presumably to be
launched from, their very central regions (\citealt{k86};
\citealt{p93}; \citealt{l97}). In addition, it now seems unlikely that
accretion discs are able to drag in the large-scale poloidal fields
assumed in the models to thread the inner disc regions
\citep{lpp94}. This has led to a number of recent suggestions that jet
acceleration, and perhaps collimation, can be due to locally
generated, small scale, perhaps tangled, magnetic fields
(\citealt{tp96}; \citealt{tbr99}; \citealt{hb00}; \citealt{kms02};
\citealt{l02}; \citealt{w02}).

Bearing this in mind, we set out here to address one specific
question, which is: are the acceleration mechanisms for the jets the
same in both AGN and YSOs? Given the uncertainties of the jet
acceleration process itself, we approach the problem in a rather
generic and abstract manner. Since we are only concerned here with the
acceleration process, and not the jet collimation, we consider the
driving of a spherically symmetric outflow by the injection of energy
into the gas at a fixed radius close to the central object\footnote{Note that we
could equivalently consider an injection of momentum rather than energy. This
might add physical reality at the expense of complexity, and here we choose to
remain with our simplified abstract approach}. This, in
general terms, must be the basis of any jet acceleration
mechanism. And, since the final jet velocities are directly related to
the size of the central object, it follows that the acceleration process
must be reasonably well localised in that vicinity. We treat the gas
in a simple manner as having a purely thermal pressure, $P$, and
internal energy, $u$, and a ratio of specific heats $\gamma$ which we
shall take to be $\gamma = 4/3$. The exact value of $\gamma$ is not
critical to our arguments, provided that $\gamma < 5/3$ so that the
outflow becomes supersonic. We note, however, that taking $\gamma =
4/3$ is in fact appropriate to the case of an optically thick
radiation-pressure dominated flow, and to the case in which the
dominant pressure within the gas is caused by a tangled magnetic field
(e.g \citealt{hb00}). Thus we feel that such treatment should
allow us to draw some general conclusions.

If the jet acceleration mechanism is the same for both YSOs and AGN,
then the same (appropriately scaled) energy input rate should account
for the jet velocities in both sets of objects. Thus, for example, we
ask whether the same energy input rate gives rise to a final jet
velocity $v_{\rm jet} \sim 2 v_{\rm esc}$ in the non-relativistic case
and $\gamma_{\rm jet} \sim 7$ in the relativistic case. With this in
mind, we undertake the following computations. In Section 2, we
consider non-relativistic outflows, relevant to YSOs. In Section 3, we
consider relativistic outflows, from compact relativistic objects,
relevant to AGN. In both cases we start from a steady configuration, in
hydrostatic equilibrium, and inject energy into the gas at a steady
rate over a small volume close to the inner radius. We follow the
time-evolution of the gas as it expands. Once the expansion has
proceeded to a large enough radius, we match the solution onto a
steady wind solution in order to estimate the final outflow
velocity. We then plot the final jet velocity as a function of the
(dimensionless) energy input rate (heating rate) for both the
relativistic and non-relativistic cases.

In Section \ref{sec:discussion}, we present our results and conclusions. 

\section{Non-relativistic (YSO) jets}
\subsection{Fluid equations}
\label{sec:nreqns}
 For YSO jets we expect the gravitational field to be well
approximated by a non-relativistic (Newtonian) description. In one
(radial) dimension the equations describing such a fluid including the
effects of energy input are expressed by the conservation of mass,
\begin{equation}
\pder{\rho}{t} + v^r\pder{\rho}{r} + \frac{\rho}{r^2}\pder{}{r}(r^2 v^r) = 0,
\label{eq:nrcty}
\end{equation}
momentum,
\begin{equation}
\pder{v^r}{t} + v^r\pder{v^r}{r} + \frac{1}{\rho}\pder{P}{r} + \frac{GM}{r^2} = 0,
\label{eq:nrmom}
\end{equation}
and energy,
\begin{equation}
\pder{(\rho u)}{t} + v^r\pder{(\rho u)}{r} + \left[\frac{P + \rho u}{r^2}\right]\pder{}{r}(r^2 v^r) = \rho\Lambda,
\label{eq:nrinten}
\end{equation}
where $\rho$, $v^r$, $P$ and $u$ are the fluid density, radial velocity,
pressure and internal energy per unit mass respectively, $M$ is the mass of the
gravitating object (in this case the central star), and
\begin{equation}
\Lambda = \frac{dQ}{dt} = T\frac{ds}{dt}
\end{equation}
is the heat energy input per unit mass per unit time (where $T$ and
$s$ are the temperature and specific entropy respectively). The
equation set is closed by the equation of state for a perfect gas in
the form
\begin{equation}
P = (\gamma - 1)\rho u.
\label{eq:nreos}
\end{equation}

\subsubsection{Scaling}
\label{sec:scaling}
To solve (\ref{eq:nrcty})-(\ref{eq:nreos}) numerically we scale the
variables in terms of a typical length, mass and timescale. These we
choose to be the inner radius of the gas reservoir $[L]=R_*$, the mass
of the gravitating body $[M] = M_*$ and the dynamical timescale at the
the inner radius ($r=R_*$), $[\tau] = (GM_*/R_*^3)^{-1/2}$. In these
units $GM=1$ and the density, pressure, velocity and internal energy,
respectively, have units of density, $[\rho] = M_*/R_*^3$, pressure,
$[P] = M_*/(R_*\tau^2)$, circular velocity at $R_*$, $[v] =
\sqrt{GM_*/R_*}$ and gravitational potential at $R_*$, $[u] =
GM_*/R_*$. Note that the net heating rate per unit mass $\Lambda$ is
therefore given in units of gravitational potential energy,
$GM_*/R_*$, per dynamical timescale at $R_*$,
$(GM_*/R_*^3)^{-1/2}$. We point out that this scaling is simply to
ensure that the numerical solution is of order unity and that when
comparing the results to the relativistic simulations we scale the
solution in terms of dimensionless variables.

\subsection{Numerical solution}
 We solve (\ref{eq:nrcty})-(\ref{eq:nreos}) in a physically intuitive
way using a staggered grid where the fluid velocity is defined on the
half grid points whereas the density, pressure, internal energy and
heating rate are specified on the integer points. This allows for
physically appropriate boundary conditions and allows us to treat the
different terms in a physical way by applying upwind differencing to
the advective terms but using centred differencing on the gradient
terms. The scheme is summarised in Figure \ref{fig:scheme} with the
discretized form of the equations given in appendix
\ref{sec:appendixA}. The staggered grid means that only three boundary
conditions are required, as shown in Figure \ref{fig:scheme}. We set
$v=0$ at the inner boundary and the density and internal energy equal
to their initial values (effectively zero) at the outer boundary.

\subsection{Initial conditions}
\label{sec:nrinit}
\begin{figure}[h]
\begin{center}
\begin{turn}{270}\epsfig{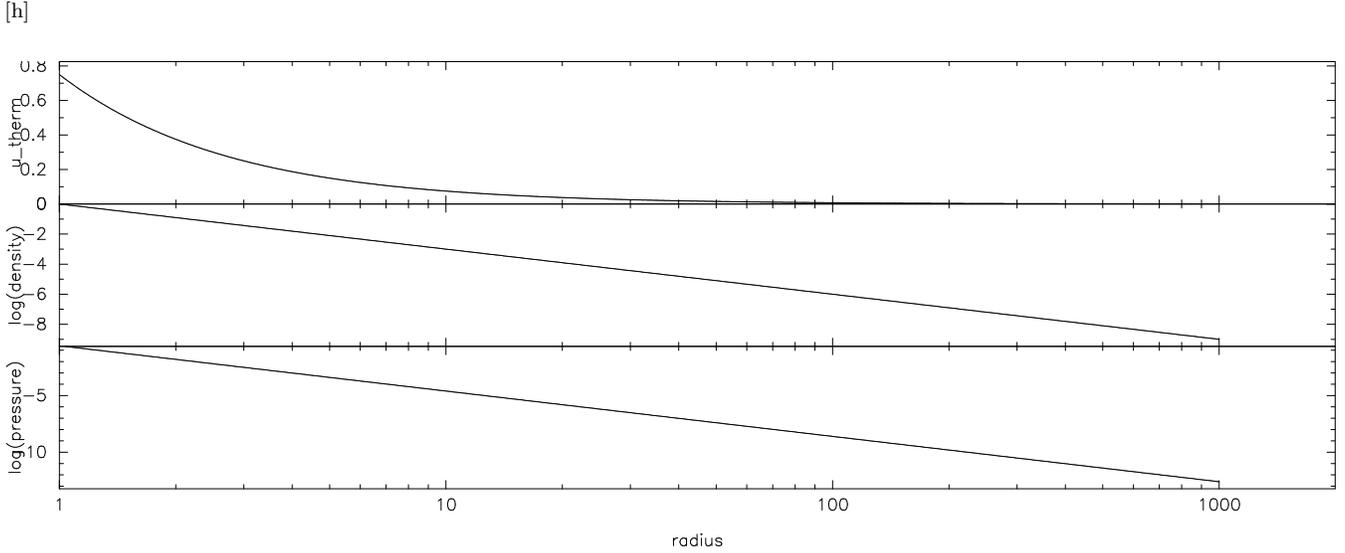}\end{turn}
\caption{The initial conditions for the non-relativistic case, We plot
profiles of density, pressure and internal energy per unit mass (or
temperature) as functions of radius. The quantities here are
dimensionless and the units are as described in \S\ref{sec:scaling}.}
\label{fig:nrinit}
\end{center}
\end{figure}

 The form of the initial conditions is not particularly crucial to the
problem, as the wind eventually reaches a quasi-steady state that is
independent of the initial setup. What the initial conditions do
affect is the time taken to reach this steady state (by determining
how much mass must initially be heated in the wind). We proceed by
setting up a body of gas (loosely `an atmosphere') above the `star'
(or rather, an unspecified source of gravity) initially in hydrostatic
equilibrium, such that $v=0$ everywhere and
\begin{equation}
\frac{dP}{dr} = -\frac{GM\rho}{r^2}.
\end{equation}
The pressure is related to the density by a polytropic equation of state
\begin{equation}
P = K\rho^\gamma,
\label{eq:polyK}
\end{equation}
where $K$ is some constant. Combining these two conditions we obtain an
equation for the density gradient as a function of radius
\begin{equation}
\frac{d\rho(r)}{dr} = -\frac{\rho(r)^{-(\gamma-2)}}{\gamma K}\frac{GM}{r^2}.
\end{equation}
Integrating this equation from $r$ to some upper bound $R_{\infty}$ we obtain
\begin{equation}
\rho(r) = \left[\frac{\gamma-1}{\gamma K}\left(\frac{GM}{r} -
\frac{GM}{R_\infty}\right)\right]^{1/(\gamma-1)}.
\label{eq:nrdrhodr}
\end{equation}
To ensure that pressure and density are finite everywhere (for
numerical stability) we set $R_{\infty}=\infty$. The density is then
given as a simple function of radius where it remains to specify the
polytropic constant $K$. In scaled units we choose
$K=(\gamma-1)/\gamma$ such that $\rho(R_*) = 1$ ({\it i.e.} the
central density equals the mean density of the gravitating body --
note that we neglect the self-gravity of the gas itself. Choosing $K$
effectively determines the amount of mass present in the atmosphere and thus the
strength of the shock front which propagates into the ambient medium (in terms
of how much mass is swept up by this front).

We set the initial pressure distribution using (\ref{eq:polyK}). If we do
this, however, the slight numerical imbalance of pressure and gravity
results in a small spurious response in the initial conditions if we
evolve the equations with zero heating. In the non-relativistic case
the spurious velocity is kept to an acceptably small level by the use
of a logarithmic radial grid (thus increasing the resolution in the
inner regions). In the relativistic case however this slight departure
from numerical hydrostatic equilibrium is more significant. This
response is therefore eliminated by solving for the pressure gradient
numerically using the same differencing that is contained in the
evolution scheme. That is we solve from the outer boundary condition
$P(r_{\rm max})=K\rho(r_{\rm max})^{\gamma}$ according to
\begin{equation}
P_{i-1} = P_i - (r_i - r_{i-1})\frac{\rho_{i-1/2}}{r^2_{i-1/2}}.
\label{eq:Pnum}
\end{equation}
Solving for the pressure in this manner reduces any spurious response
in the initial conditions to below round--off error. The internal
energy is then given from (\ref{eq:nreos}). The pressure calculated
using (\ref{eq:Pnum}) is essentially indistinguishable from that found
using (\ref{eq:polyK})($\Delta P/P\sim 10^{-5})$. The initial
conditions calculated using equation~(\ref{eq:nrdrhodr}),
(\ref{eq:Pnum}) and (\ref{eq:nreos}) are shown in Figure
\ref{fig:nrinit}. We use a logarithmic grid with 1001 radial grid
points, setting the outer boundary at $r/R_* = 10^3$. Using a higher spatial resolution
does not affect the simulation results.

\subsubsection{Heating profile}
 The choice of the shape of the heating profile $\Lambda(r)$ is fairly
arbitrary since we wish simply to make a comparison between the
non-relativistic and relativistic results. We choose to heat the wind in a
spherical shell of a fixed width using a linearly increasing and then
decreasing heating rate, symmetric about some heating radius $r_{\rm heat}$
which we place at $r=2.1 R_*$. The heating profile is spread over a radial zone
of width  $2R_*$ (that is the heating zone extends from $r=1.1 R_*$ to $3.1
R_*$)(see Figure \ref{fig:nrresults}). We choose a heating profile of this form
such that it is narrow enough to be associated with a particular radius of
heating (necessary since we are looking for scaling laws) whilst being wide
enough to avoid the need for high spatial resolution or complicated algorithms (necessary if the heat
input zone is too narrow). The important parameter is thus the \emph{location}
of the heating with respect to the Schwarzschild radius, so long as the heating
profiles are the same in both the relativistic and non-relativistic cases.
Provided that the heating profile is narrow enough to be associated with a
particular radius and wide enough to avoid numerical problems, our results do
not depend on the actual shape of the profile we choose.

\subsection{Results}
\label{sec:nrresults}
\begin{figure}[t]
\begin{turn}{270}\epsfig{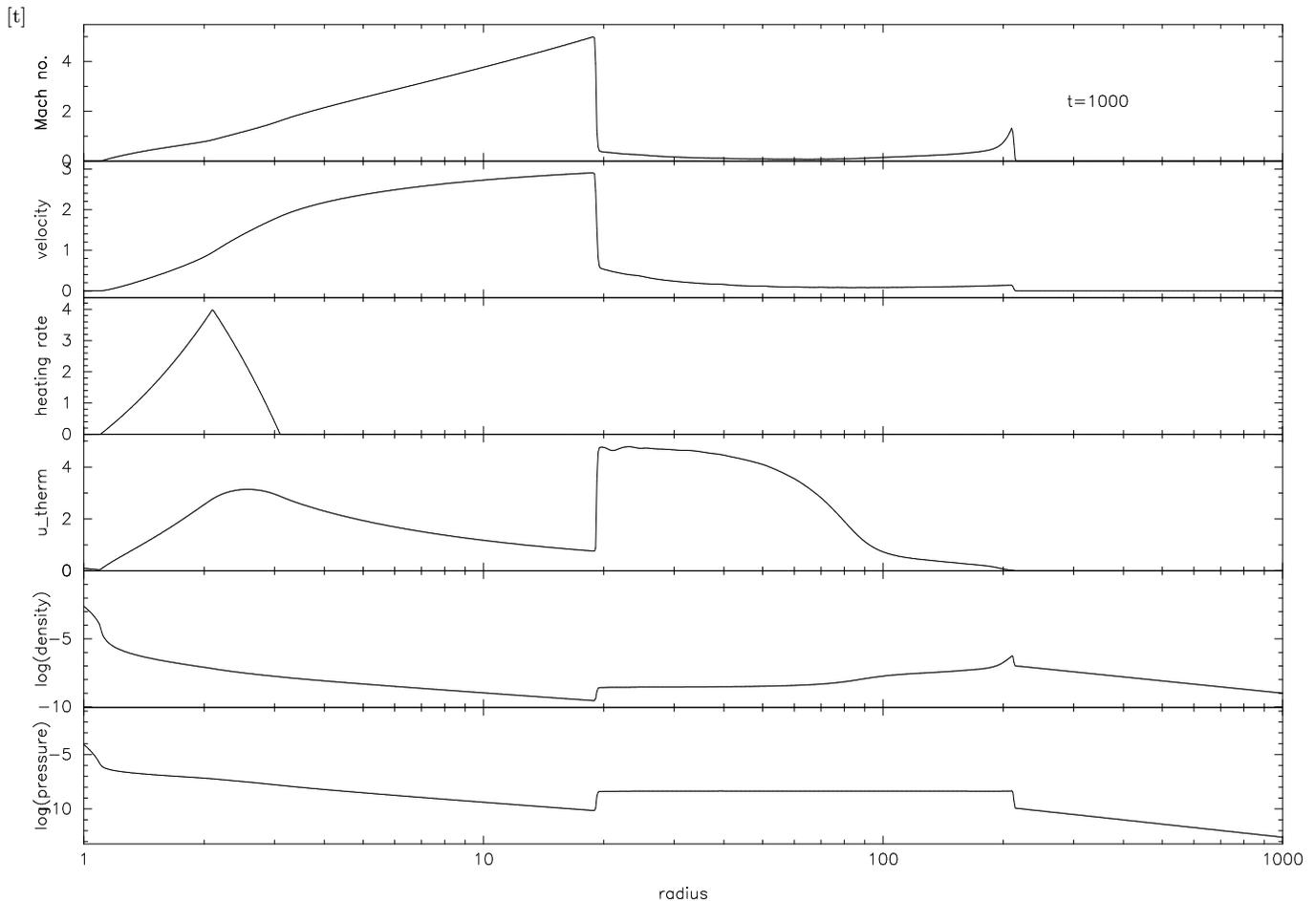}\end{turn}
\caption{Results of a typical non-relativistic simulation at time $t=
1000$ (where units of time are the dynamical time at the innermost
radius, $\sqrt{R_*^3/GM}$).  Quantities shown are the Mach number
($v/c_s$), velocity, heating rate ($\Lambda$), internal energy per
unit mass ($u \equiv u_{\rm therm}$), log(density) and log(pressure).}
\label{fig:nrresults}
\end{figure}

\begin{figure}
\begin{center}
\begin{turn}{270}\epsfig{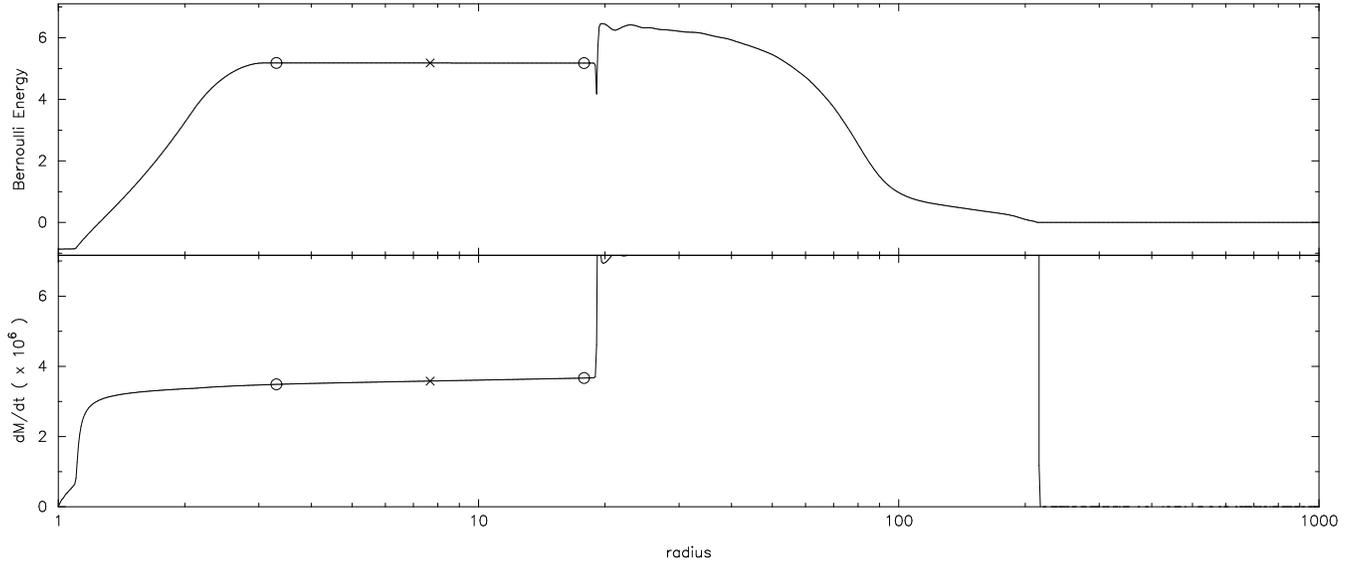}\end{turn}
\caption{Bernoulli energy $E=\frac{1}{2}v^2 + \rho u + P -
GM/r$ (top) and mass outflow rate $\dot{M} = 4\pi r^2\rho v$
(bottom) in the time-dependent wind solution at $t= 1000$. The
profiles are approximately constant over the region between the two
circles. The sample point used to match this flow to the appropriate
steady state solution is indicated by a cross.}
\label{fig:nrmdoten}
\end{center}
\end{figure}

 The results of a typical non-relativistic simulation with a moderate
heating rate are shown in Figure \ref{fig:nrresults} at $t=1001$
(where $t$ has units of the dynamical time at the inner radius).
We observe the effect of the heating propagating outwards in the
atmosphere in the form of a shock front. After several hundred
dynamical times the wind structure approaches a steady state in that
there is only a small change of the overall wind structure due to the
shock continuing to propagate outwards into the surrounding
medium. The small disturbance propagating well ahead of the main shock
is a transient resulting from the response of the atmosphere to the
instantaneous switch-on of the heating. The velocity of the gas begins to asymptote to a constant
value as the shock propagates outwards.  Plotting the mass outflow
rate $\dot{M} = 4\pi r^2\rho v$ and the Bernoulli energy $E =
\frac{1}{2}v^2 + \rho u + P - GM/r$ as a function of radius
(Figure \ref{fig:nrmdoten}),
we see that indeed the wind structure is eventually close to that of a
steady wind above the heating zone (ie. $\dot{M}$ and $E \sim$
constant). It is thus computationally inefficient and impractical to
compute the time-dependent solution for long enough to determine an
accurate velocity as $r \rightarrow \infty$ when the wind will continue to
have a steady structure. Instead we find the steady wind solution for
a given amount of energy input to the wind corresponding to the energy
plotted in Figure \ref{fig:nrmdoten} (top panel).

\subsection{Steady wind solution}
\begin{figure}[t]
\begin{center}
\begin{turn}{270}\epsfig{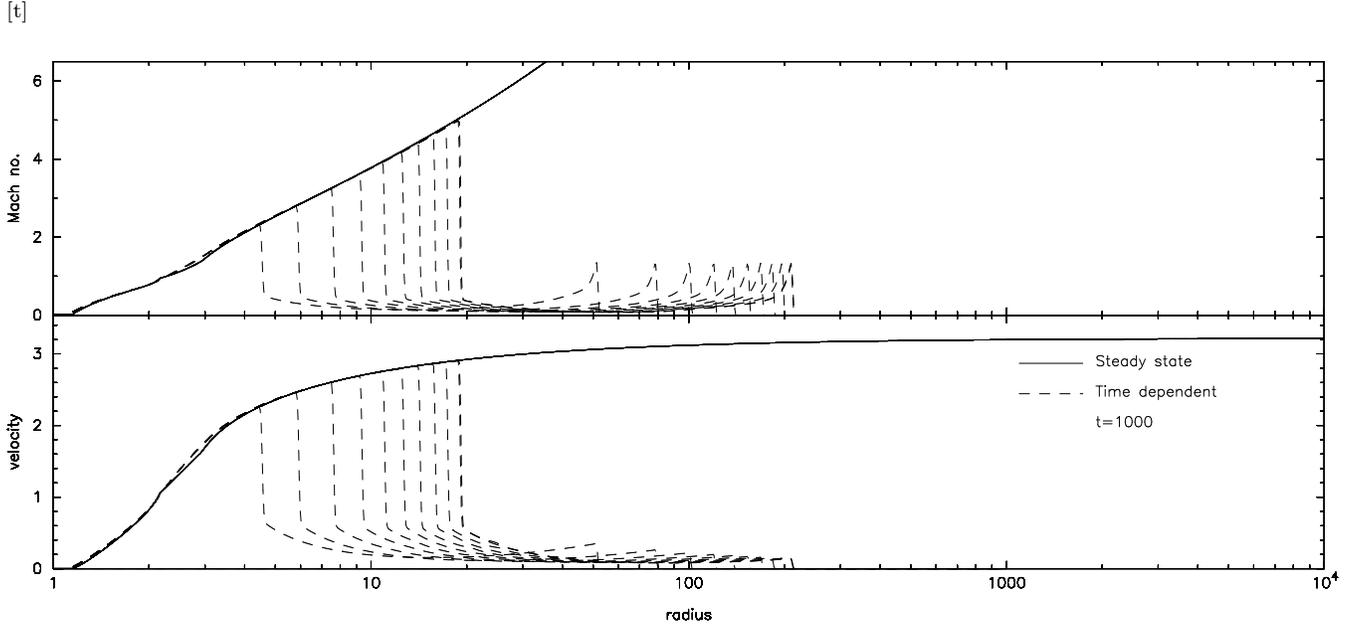}\end{turn}
\caption{Steady wind Mach number (top panel) and velocity (centre
panel) profiles are compared to the time-dependent solution (plotted
every 100 dynamical times). There is a small discrepancy between the two
solutions where we have taken the limit in approaching the singular point at $M=1$, but an otherwise excellent agreement between the two
solutions.}
\label{fig:nrsteadysoln}
\end{center}
\end{figure}

Non-relativistic, steady state ($\partial/\partial t = 0$) winds with
energy input have been well studied by many authors, and the equations
describing them can be found in \citet{lc99}, who credit the original
work to \citet{ha70}. The reader is thus referred to \citet{lc99} for
details of the derivation. As in the usual Bondi/Parker
(\citealt{b52}, \citealt{p58}) wind solution with no heat input, we
set $\partial/\partial t = 0$ in (\ref{eq:nrcty})-(\ref{eq:nreos}) and
combine these equations into one equation for the Mach number $M^2 =
v^2/c_s^2$ as a function of radius, given by
\begin{equation}
\frac{dM^2}{dr} = -\frac{M^2(2+(\gamma-1)M^2)}{2(M^2-1)\left[e(r) + GM/r\right]}
  \left[(1+M^2\gamma)\frac{dQ}{dr} +
\frac{GM}{r^2}\frac{(5-3\gamma)}{(\gamma-1)} - \frac{4e(r)}{r}\right],
\label{eq:machno}
\end{equation}
where $dQ/dr$ is the local heating gradient and $e(r)$ is the Bernoulli energy which is specified by integrating the Bernoulli equation
\begin{equation}
\frac{de(r)}{dr} = \frac{d}{dr}\left[\frac{1}{2}v^2 + \rho u + P - \frac{GM}{r}\right] = \frac{dQ}{dr}, 
\end{equation}
to give
\begin{eqnarray}
e(r) & = & e(r_\infty) - Q(r) \nonumber \\
& = & e(r_\infty) - \int^{r_\infty}_r\frac{dQ}{dr},
\label{eq:qsubtract}
\end{eqnarray}
where $Q(r)$ is the total energy input to the wind. Since we are interested in
the terminal velocity of the outflow we choose a point above the heating shell
where the energy has reached its steady state value ({\it i.e.} where the energy is constant in Figure
\ref{fig:nrmdoten}, top panel) and integrate outwards using
the energy and Mach number at this point to solve (\ref{eq:machno}) as an initial value
problem. Note that in fact the terminal velocity is determined by
the (constant) value of the Bernoulli energy above the heating zone since as $r\to\infty$, $e(r)\to \frac{1}{2}v^2$. However we
compute the steady wind profiles both inwards and outwards to show the
consistency between the time-dependent solution and the steady state version.

 In order to perform the inward integration, we must determine the energy at
every point for our steady solution by subtracting the heat input from
the steady state energy as we integrate inwards through the heating shell
($\ref{eq:qsubtract}$). To determine this however we must also determine the local
(steady state) heating gradient $dQ/dr$, which is related to the (time dependent) heating rate
$\Lambda$ by setting $\partial/\partial t=0$ in the time dependent
version, ie.
\begin{equation}
\Lambda = \frac{dQ}{dt} = \pder{Q}{t} + v\frac{dQ}{dr} = v\frac{dQ}{dr}.
\end{equation}
 We therefore calculate $dQ/dr$ from the time dependent solution using
\begin{equation}
\frac{dQ}{dr} = \frac{\Lambda(r)}{v(r)},
\end{equation}
where $v(r)$ is the wind velocity at each point in the heating shell
from the time-dependent solution. The problem with this is that at the
inner edge of the heating shell the heating rate is finite while the
velocity is very close to zero, resulting in a slight overestimate of
the total energy input near the inner edge of the shell in the steady
wind solution. Care must also be taken in integrating through the singular point
in equation (\ref{eq:machno}) at $M^2=1$.  Most authors
(e.g. \citealt{lc99}) solve the steady wind equations starting from
this point but for our purposes it is better to start the
integration outside of the heating shell where the energy is well
determined. We integrate through the critical point by using a
first order Taylor expansion and appropriate limit(s), although
this introduces a small discrepancy between the steady state and time-dependent
results in this region (Figure \ref{fig:nrsteadysoln}).

 Having determined the energy and heating gradient at each point in the wind we
integrate (\ref{eq:machno}) both inwards and outwards from the chosen point
above the heating shell using a
fourth order Runge-Kutta integrator (scaling (\ref{eq:machno}) to the
units described in \S\ref{sec:scaling}). The velocity profile is then
given by $v^2 = M^2 c_s^2$ where
\begin{equation}
c_s^2(r) = \frac{2(\gamma-1)}{2 + M^2(r)(\gamma-1)}\left[e(r) + \frac{GM}{r}\right]. 
\end{equation}
 The resulting steady wind solution is shown in Figure
\ref{fig:nrsteadysoln} along with the time-dependent solution. The two
profiles are in excellent agreement, proving the validity of our
time-dependent numerical solution and the assumption that the wind is in a steady
state. The steady solution thus provides an accurate estimate of the velocity at
arbitrarily large radii (although as pointed out previously this is set by the
value of the steady state Bernoulli energy).

\subsection{Terminal wind velocities as a function of heating rate}
\label{sec:nrrates}
 Using the steady wind extrapolation of the time-dependent solution,
we can determine the relationship between the heating rate and the
terminal wind velocities. In order to make a useful comparison between
the heating rates used in both the Newtonian and the relativistic
regimes, we need to define a local canonical heating rate
$\Lambda_c(r)$ valid in both sets of regimes. In dimensional terms the
heating rate $\Lambda(r)$ corresponds to an input energy per unit mass
per unit time. Thus we need to define the local canonical heating rate
as

\begin{equation}
\Lambda_c(r) = \frac{\Delta E}{\Delta t},
\end{equation}
for some relevant energy $\Delta E$ and some relevant timescale
$\Delta t$.

 There are clearly many different ways in which we might define a
canonical heating rate. We find, however, that our results are not
sensitive to the particular choice we make. We shall make use of a
definition which draws on the physical processes we expect to be
behind the jet acceleration process. Even though the processes by
which this occurs are still obscure, we expect the energy for the jet
to be provided fundamentally by liberation of energy in a rotating
flow. Thus, with this physical motivation in mind, we take the
canonical energy per unit mass, $\Delta E$, to be the energy released
locally by bringing to rest a particle of unit mass which is orbiting
in a circular orbit at radius $r$.  In the Newtonian regime this is
simply the kinetic energy of a circular orbit
\begin{equation}
\Delta E = \frac{1}{2}v_{\phi}^2 = \frac{GM}{2r}.
\end{equation}
(An alternative possibility, for example, would be to take $\Delta E$
to be the energy released by dropping a particle from infinity and
bringing it to rest at radius $r$, which would correspond to the
escape energy from that radius, $GM/r$.) By similar reasoning, we take
the canonical timescale on which the energy is released to be the
orbital timescale at radius $r$, that is $\Delta t = \Omega_o^{-1}$,
where
\begin{equation}
\Omega_o = (GM/r^3)^{1/2}.
\end{equation}
Using this, we are now in a position to define a local canonical
heating rate as
\begin{equation}
\Lambda_c(r) = \Delta E  \times \Omega_o = \frac{(GM)^{3/2}}{2 r^{5/2}}.
\label{eq:nrlambdac}
\end{equation}
For intercomparison of our various wind computations both in the
Newtonian and in the relativistic regimes, we now use the canonical
heating rate derived above to define a dimensionless heating rate for
each wind computation. Because heat is added over a range of radii, we
need to define the dimensionless heating rate $\langle \Lambda
\rangle$ as an appropriate volume average. We shall define

\begin{equation}
\langle \Lambda \rangle = \frac{\int^{r_2}_{r_1}\Lambda r^2
dr}{\Lambda_c(r_{\rm max}) \int^{r_2}_{r_1} r^2 dr},
\label{eq:lambdaav}
\end{equation}
where $r_{\rm max}$ is the radius at which the heating rate
$\Lambda(r)$ takes its maximum value and $r_1$ and $r_2$ are the lower
and upper bounds of the heating shell respectively.

\begin{figure}[t]
\begin{center}
\begin{turn}{270}\epsfig{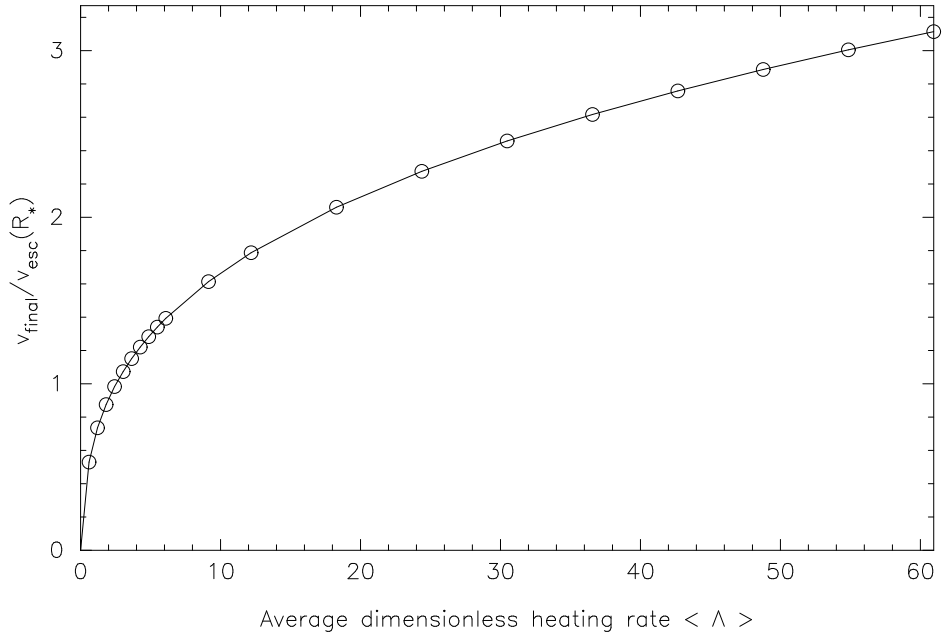}\end{turn}
\caption{Terminal wind velocities plotted as a function of the average
dimensionless heating rate $\langle \Lambda \rangle$. Wind velocities
are plotted in units of the escape velocity at the inner radius
(ie. $r=R_*=1$), $v_{\rm esc} = (2GM/R_*)^{1/2}$. We compute solutions
corresponding to velocities typically observed in YSO jets (with a
fairly generous upper limit of $v/v_{\rm esc}\sim 3$).}
\label{fig:nrrates}
\end{center}
\end{figure}

 The relation between this average dimensionless heating rate and the
terminal wind velocity is shown in Figure \ref{fig:nrrates}. The
wind velocities are plotted in units of the escape velocity $v_{\rm
esc}$ at $R_*$ and solutions are computed for wind velocities of up to
$\sim 3v_{\rm esc}$. The important point in the present analysis is that
the heating rate can be meaningfully compared to the relativistic
results (see below).

\section{Relativistic jets}
 Having determined the heating rates required to produce the observed
velocities in YSO jets we wish to perform exactly the same calculation
within a relativistic framework. We proceed in precisely the same
manner as in the non-relativistic case. We adopt the usual convention
that Greek indices run over the four dimensions 0,1,2,3 while Latin
indices run over the three spatial dimensions 1,2,3. Repeated indices
imply a summation and a semicolon refers to the covariant derivative.
The density $\rho$ refers to the rest mass density only, that is
$\rho=nm_{0}$ where $n$ is the number density of baryons and $m_0$ is
the mass per baryon.

\subsection{Fluid equations}
 The equations describing a relativistic fluid are derived from the conservation
 of baryon number,
\begin{equation}
(\rho U^\mu)_{;\mu} = 0,
\label{eq:relcty1}
\end{equation}
the conservation of energy-momentum projected along a direction
perpendicular to the four velocity $U^\mu$ (which gives the equation of
motion),
\begin{equation}
h_{\mu\alpha}T^{\alpha\nu}_{\phantom{\alpha\nu};\nu} = (g_{\mu\alpha} + U_{\mu}U_{\alpha})T^{\alpha\nu}_{\phantom{\alpha\nu};\nu} = 0,
\label{eq:relmom1}
\end{equation}
and projected in the direction of the four-velocity (which gives the
energy equation),
\begin{equation}
U_{\alpha}T^{\alpha\nu}_{\phantom{\alpha\nu};\nu} = 0.
\label{eq:relen}
\end{equation}
Here the quantity $T^{\mu\nu}$ is the energy momentum tensor, which
for a perfect fluid is given by
\begin{equation}
c^2 T^{\mu\nu} = \rho h U^{\mu}U^{\nu} + Pg^{\mu\nu},
\label{eq:tuv}
\end{equation}
where $h$ is the specific enthalpy,
\begin{equation}
h = c^2 + u + \frac{P}{\rho} = c^2 + \frac{\gamma P}{(\gamma-1)\rho}.
\end{equation}
As in the non-relativistic case $u$ is the internal energy per unit
mass, $P$ is the gas pressure and we have used the equation of state
given by equation~(\ref{eq:nreos}). The energy equation may also be
derived from the first law of thermodynamics using
equation~(\ref{eq:relcty1}), which is a more convenient way of
deriving an energy equation in terms of the internal energy (rather
than the total energy) and in this case ensures that the meaning of
the heating term is clear. The metric tensor is given by the
Schwarzschild (exterior) solution to Einstein's equations, that is
\begin{equation}
ds^2 = -c^2d\tau^2 = -\left(1 - \frac{2GM}{c^2 r}\right) c^2dt^2 + \left(1 - \frac{2GM}{c^2
r}\right)^{-1} dr^2 + r^2 (d\theta^2 + \mathrm{sin}^2\theta d\phi^2).
\label{eq:metric}
\end{equation} 
We consider radial flow such that $U^\theta = U^\phi = 0$.
The four velocity is normalised such that
\begin{equation}
U_\mu U^\mu = -c^2,
\end{equation}
and we define
\begin{equation}
U^t \equiv \frac{dt}{d\tau} = \left(1 - \frac{2GM}{c^2 r}\right)^{-1}\left[\left(1 -
\frac{2GM}{c^2 r}\right) + \frac{(U^r)^2}{c^2}\right]^{1/2},
\label{eq:u0}
\end{equation}
which we denote as
\begin{equation}
U^t = \frac{\Gamma}{\alpha^2}
\end{equation}
where we set for convenience
\begin{equation}
\Gamma = \left[\left(1 - \frac{2GM}{c^2 r}\right) + \frac{(U^r)^2}{c^2}\right]^{1/2},
\end{equation}
and
\begin{equation}
\alpha^2 = \left(1 -\frac{2GM}{c^2 r}\right).
\end{equation} 
 Note that while $\alpha$ corresponds to the lapse function in the
$3+1$ formulation of general relativity, the quantity $\Gamma$ is
\emph{not} the Lorentz factor of the gas (which we denote as $W$) as it
is usually defined in numerical relativity (e.g. \citealt{bea97}) but
is related to it by $W = \Gamma/\alpha$.  From (\ref{eq:u0}) we also
have the relation
\begin{equation}
\pder{U^t}{t} = \frac{U^r}{\alpha^2\Gamma c^2}\pder{U^r}{t}
\label{eq:du0dt}
\end{equation}

 From (\ref{eq:relcty1}), (\ref{eq:relmom1}) and (\ref{eq:relen}) using
(\ref{eq:tuv}), (\ref{eq:metric}), (\ref{eq:u0}) and (\ref{eq:du0dt}) we thus derive the
continuity equation,
\begin{equation}
\pder{\rho}{t} + v^r\pder{\rho}{r} + \frac{\alpha^2
\rho}{\Gamma}\left[\frac{1}{r^2}\pder{}{r}(r^2 U^r) + \frac{U^r}{\alpha^2\Gamma c^2}\pder{U^r}{t}\right] = 0, \label{eq:relcty}
\end{equation}
the equation of motion,
\begin{equation}
\pder{U^r}{t} + v^r\pder{U^r}{r} + \frac{\Gamma \alpha^2 c^2}{\rho h}\pder{P}{r} +
\frac{U^r}{\rho h}\pder{P}{t} + \frac{\alpha^2}{\Gamma}\frac{GM}{r^2} = 0,
\label{eq:relmom}
\end{equation}
and the internal energy equation,
\begin{equation}
\pder{(\rho u)}{t} + v^r\pder{(\rho u)}{r} +
\frac{\alpha^2}{\Gamma}\left(P + \rho
u\right)\left[\frac{1}{r^2}\pder{}{r}(r^2 U^r) + \frac{U^r}{\alpha^2\Gamma
c^2}\pder{U^r}{t}\right] = \frac{\alpha^2}{\Gamma}\rho\Lambda,
\label{eq:relinten}
\end{equation}
where
\begin{equation}
v^r \equiv \frac{U^r}{U^t} \equiv \frac{dr}{dt}
\end{equation}
is the velocity in the co-ordinate basis. We define the heating rate
per unit mass as
\begin{equation}
\Lambda \equiv T\frac{ds}{d\tau},
\end{equation}
where $T$ is the temperature, $s$ is the specific entropy and $d\tau$
refers to the local proper time interval ($\Lambda$ is therefore a
local rate of energy input, caused by local physics). A comparison of
(\ref{eq:relcty}), (\ref{eq:relmom}) and (\ref{eq:relinten}) with their
non-relativistic counterparts (\ref{eq:nrcty}), (\ref{eq:nrmom}) and
(\ref{eq:nrinten}) shows that they reduce to the non-relativistic
expressions in the limit as $c\to \infty$, and to special relativity
as $M\to 0$.

The `source terms' containing time derivatives of $U^r$ and $P$ are
then eliminated between the three equations using the equation of
state (\ref{eq:nreos}) to relate pressure and internal
energy. Substituting for pressure in (\ref{eq:relinten}) and
substituting this into (\ref{eq:relmom}) we obtain the equation of
motion in terms of known variables,
\begin{equation}
\pder{U^r}{t} + \frac{v^r}{X}\left(1-\frac{\gamma P}{\rho h}\right)\pder{U^r}{r}
= - \frac{c^2 \alpha^4}{\rho h \Gamma X}\pder{P}{r} - \frac{\alpha^2}{\Gamma
X}\frac{GM}{r^2} + \frac{v^r}{X}\frac{\gamma P}{\rho h}\frac{2U^r}{r} -
\frac{v^r}{hX}(\gamma-1)\Lambda,
\label{eq:relmomuse}
\end{equation}
where for convenience we define
\begin{equation}
X \equiv 1 - \left(\frac{\gamma P}{\rho h}\right)\frac{U^r U^r}{\Gamma^2 c^2},
\end{equation}
and we have expanded the $\frac{1}{r^2}\pder{}{r}(r^2
U^r)$ terms in order to combine the spatial derivatives of $U^r$ into
one term. We then substitute (\ref{eq:relmomuse}) into
(\ref{eq:relcty}) and (\ref{eq:relinten}) to obtain equations for the
density
\begin{equation}
\pder{\rho}{t} + v^r\pder{\rho}{r} = -\frac{\alpha^2}{\Gamma}\left[\rho A -
\frac{v^r}{h\Gamma X}\pder{P}{r} -
 \frac{U^r U^r}{\Gamma^2 c^2 X}\frac{(\gamma-1)}{h}\rho\Lambda\right], 
\label{eq:relctyuse}
\end{equation}
and internal energy,
\begin{equation}
\pder{(\rho u)}{t} + v^r\left(1-\frac{\gamma P}{\rho h}\frac{\alpha^2}{\Gamma^2
X}\right)\pder{(\rho u)}{r} = - \frac{\alpha^2}{\Gamma}\left[(P + \rho
u)A - \left(1 + \frac{U^r U^r}{\Gamma^2 c^2 X}\frac{\gamma
P}{\rho h}\right)\rho\Lambda \right].
\end{equation}
where for convenience we have defined
\begin{equation}
A \equiv \left[1-\frac{U^r U^r}{\Gamma^2 c^2 X}\left(1-\frac{\gamma P}{\rho h}\right)\right]\pder{U^r}{r}
 + \left[1 + \frac{U^r U^r}{\Gamma^2 c^2 X}\left(\frac{\gamma P}{\rho
 h}\right)\right]\frac{2U^r}{r} - \frac{U^r}{\Gamma^2 c^2 X}\frac{GM}{r^2}
\end{equation}

 From the solution specifying $U^r$ we calculate the velocity measured by an
observer at rest with respect to the time slice (referred to as \emph{Eulerian}
observers), which is given by
\begin{equation}
\bar{v}^r = \frac{U^r}{\alpha U^t} = \frac{v^r}{\alpha},
\label{eq:veul}
\end{equation}
since there are no off-diagonal terms (ie. zero shift vector) in the
Schwarzschild solution. For these observers the Lorentz factor is
given by
\begin{equation}
W = \left(1-\frac{\bar{v}^r \bar{v}_r}{c^2}\right)^{-1/2},
\label{eq:Lfac}
\end{equation}
where $\bar{v}^r \bar{v}_r = g_{rr}\bar{v}^r \bar{v}^r$, such that $U^r = W
\bar{v}^r$.

\subsection{Scaling}
\label{sec:relscaling}
 The usual practice in numerical relativity is to scale in so-called
geometric units such that $G=M=c=1$. In these units the length scale
would be the geometric radius $GM/c^2$ and the velocity would have
units of $c$. Instead for the current problem, we adopt a scaling
analogous to that of the non-relativistic case, that is we choose the
length scale to be the radius of the central object, $R_*$, where
$R_*$ is given as some multiple of the geometric radius, ie.
\begin{equation}
[L] = R_* = n\frac{GM_*}{c^2},
\end{equation}
with $n \ge 2.0$.  The mass scale is again the central object mass $[M]
 = M_*$ and the timescale is given by
\begin{equation}
[\tau] = \left(\frac{GM_*}{R_*^3}\right)^{-1/2} = n^{3/2}\frac{GM_*}{c^3}
\end{equation}
 In these units, velocity is measured in units of $[v] = n^{-1/2} c$
(or equivalently $c^2 = n$). The scaled equations are thus given
simply by setting $G=M=1$ and $c^2=n$ everywhere.

 This scaling ensures that the relativistic terms tend to zero when
$c$ (or $n$) is large and that the numerical values of $\rho$, $\rho
u$ and $U^r$ are of order unity. We thus specify the degree to which
the gravity/gas dynamics is relativistic by specifying the value of
$n$ ({\it i.e.} the proximity of the innermost radius, and thus the
heating, to the Schwarzschild radius, $R_{\rm Sch} = 2GM/c^2$). We
compute solutions corresponding to gas very close to a black hole
(highly relativistic, $n=2.0$, or $R_* = R_{\rm Sch}$), neutron star
(moderately relativistic, $n=5$, or $R_* = R_{\rm NS} = 5GM/c^2$,
which is equivalent to heating further out and over a wider region
around a black hole) and white dwarf/non-relativistic star
(essentially non-relativistic, $n=5000$, or $R_* = 2500 R_{\rm
Sch}$). Note that in the highly relativistic case although we scale
the solution to $n=2.0$ such that the mass, length and time scales
(and therefore the units of heating rate, energy etc.)  correspond to
those at $r=R_{\rm Sch}$, our numerical grid cannot begin at $R_*$ as
it does in the other cases. We therefore set the lower bound on the
radial grid to slightly below the heating shell (typically $r =
1.01R_*$ where the heating begins at $1.1R_*$).  Note that the above
scaling is merely to ensure that the numerical solution is of order
unity, since we scale in terms of dimensionless variables to compare
with the non-relativistic solution.

\subsection{Numerical Solution}
 In order to solve the relativistic fluid equations numerically we use a
method analogous to that used in the non-relativistic case (Figure
\ref{fig:scheme}). That is, we first compute $U^r$ on the staggered (half) grid
and use this to solve for $\rho$ and $\rho u$ on the integer grid points. Again
the advective terms are discretized using upwind differences (where the
`upwindedness' is determined from the sign of the co-ordinate velocity $v^r$)
and other derivatives are calculated using centred differences. As in the
non-relativistic case, where a centred difference is used, the quantities
multiplying the derivative are interpolated onto the half grid points if
necessary. In equation (\ref{eq:relctyuse}) we evaluate the $\partial
P/\partial r$ term using upwind differences.

\subsection{Initial Conditions}
\begin{figure}[t]
\begin{center}
\begin{turn}{270}\epsfig{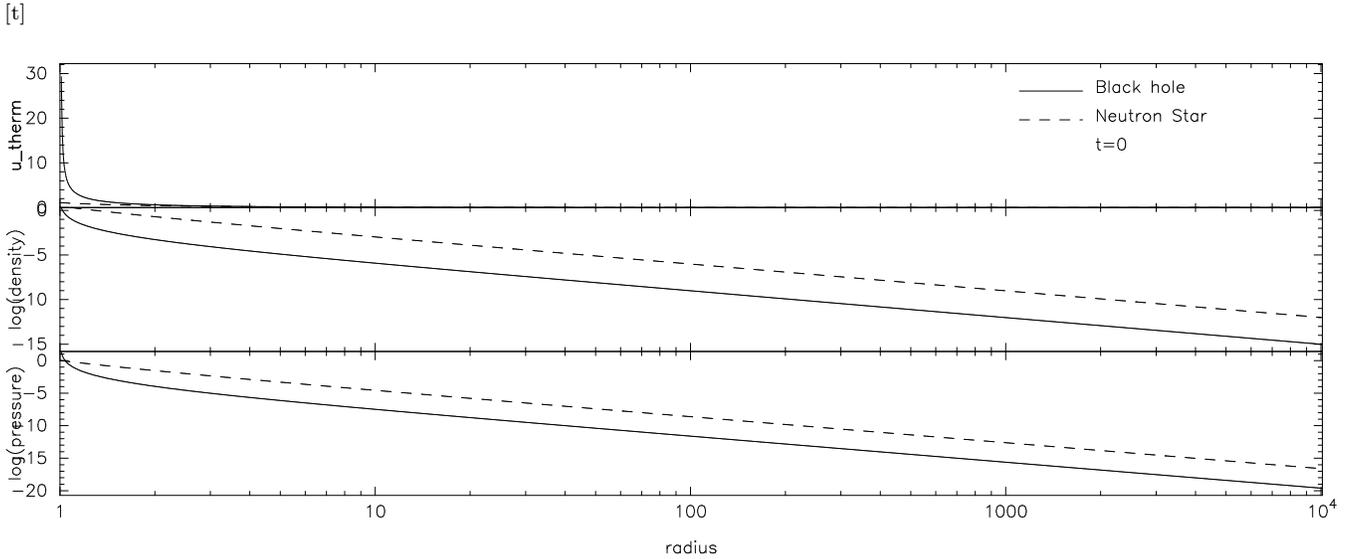}\end{turn}
\caption{The initial conditions for the gas reservoir for the
relativistic cases of a neutron star (dashed line) ($R_*/R_{\rm
Sch}=2.5$) and black hole (solid line)($R_*/R_{\rm Sch}=1.0$). Note,
however, that the innermost radius is at $r=1.01R_*$ in the latter
case.  We plot profiles of internal energy per unit mass (or
temperature), density and pressure, as functions of radius. These
quantities are given in units of $GM/R_*$, $M/R_*^3$ and $M_*/(R_*
t_*^2)$ respectively. Note that steeper gradients are required to hold the
gas in hydrostatic equilibrium as the gravitational field becomes more
relativistic. The black hole reservoir is of lower density than the neutron star version
because of the choice of the polytropic constant (chosen such that the central
density is of order unity).}
\label{fig:relinit}
\end{center}
\end{figure}

 We determine initial conditions for the relativistic case by setting
$U^r = 0$ and $\partial/\partial t=0$ in (\ref{eq:relmomuse}), from
which we have
\begin{equation}
\frac{dP}{dr} = -\frac{\rho h}{c^2}\frac{GM}{r^2}\left(1-\frac{2GM}{c^2 r}\right)^{-1}.
\label{eq:reldPdr}
\end{equation}
 The pressure is thus calculated as a function of $\rho$, $u$ and $P$
(where $P = (\gamma-1)\rho u$). We solve (\ref{eq:reldPdr}) using the
same assumptions as in the non-relativistic case (\S\ref{sec:nrinit}),
that is an adiabatic atmosphere such that
\begin{equation}
P=K\rho^\gamma.
\label{eq:relpadiabatic}
\end{equation}
We therefore have
\begin{equation}
\frac{d\rho}{dr} = -\frac{1}{\gamma K\alpha^2}\left[\rho^{(2-\gamma)} + \frac{\gamma K
\rho}{c^2(\gamma-1)}\right] \frac{GM}{r^2},
\end{equation}
which we solve using a first order (Euler) discretization to obtain a
density profile. The pressure may then be calculated using
(\ref{eq:relpadiabatic}), however to ensure that hydrostatic
equilibrium is enforced numerically we solve (\ref{eq:reldPdr}) using
the same discretization as in the fluid equations, integrating inwards
from the outer boundary condition
$P(r_{\rm max})=K\rho(r_{\rm max})^{\gamma}$. However in this case the
pressure gradient also depends on the pressure, so we use the pressure
calculated from (\ref{eq:relpadiabatic}) to calculate the initial
value of the specific enthalpy $h$ and iterate the solution until
converged ($[P^{n+1}-P^n]/P^n < 10^{-10}$). In the black hole case the
resulting pressure differs from that found using
(\ref{eq:relpadiabatic}) by $\Delta P/P \sim 10^{-2}$. We choose $K$
such that the central density is of order unity -- typically we use $K =
10\gamma/(\gamma-1)$ in the black hole case. Note that changing $K$
simply changes the amount of matter present in the atmosphere but does
not affect the temperature scaling and does not affect the final
results (although it significantly affects the integration time since it
determines the strength of the shock front and the amount of mass to be
accelerated).

 Initial conditions calculated in this manner for the black hole
($R_*/R_{\rm Sch} = n/2 =1.0$) and neutron star ($R_*/R_{\rm Sch}=2.5$)
atmospheres are shown in Figure \ref{fig:relinit}. The initial setup reduces
to that of Figure \ref{fig:nrinit} in the non-relativistic limit when the same
value of $K$ is used. We set the outer boundary at $r/R_* = 10^4$,
using 1335 radial grid points (again on a logarithmic grid).

\subsection{Results}
\label{sec:relresults}
\begin{figure}[t]
\begin{turn}{270}\epsfig{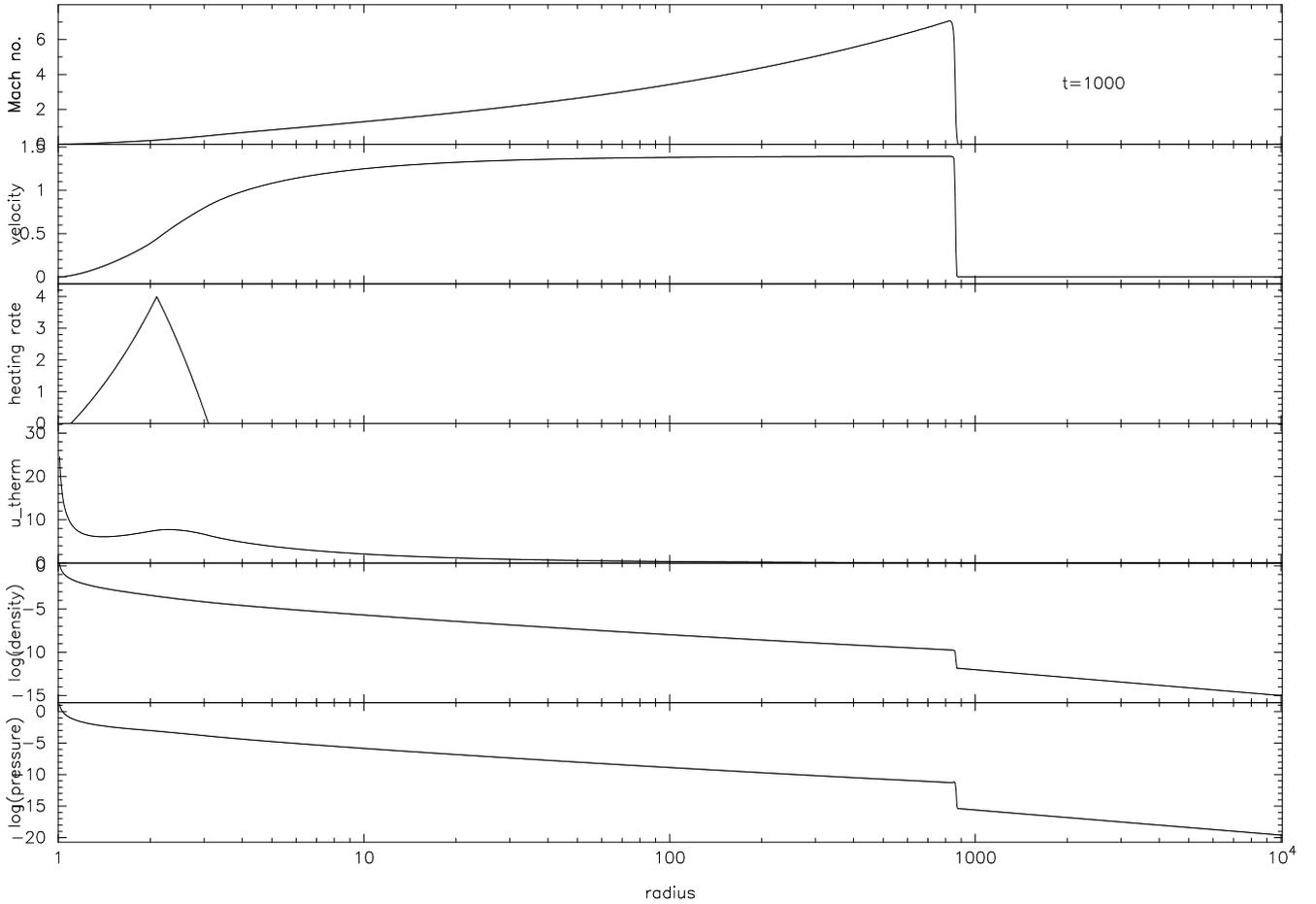}\end{turn}
\caption{Results of a typical black hole relativistic simulation at
t=1000 (where units of time are the dynamical time at the central
object). Quantities shown are the Mach number ($v/c_s$), velocity for
Eulerian observers ($\bar{v}^r$), heating rate ($\Lambda$), internal
energy per unit mass ($u \equiv u_{\rm therm}$), log(density) and
log(pressure). Units of velocity are such that $c = \sqrt{2}$ and as
in the non-relativistic case energy has units of $GM/R_*$.}
\label{fig:relresults}
\end{figure}

\begin{figure}[t]
\begin{center}
\begin{turn}{270}\epsfig{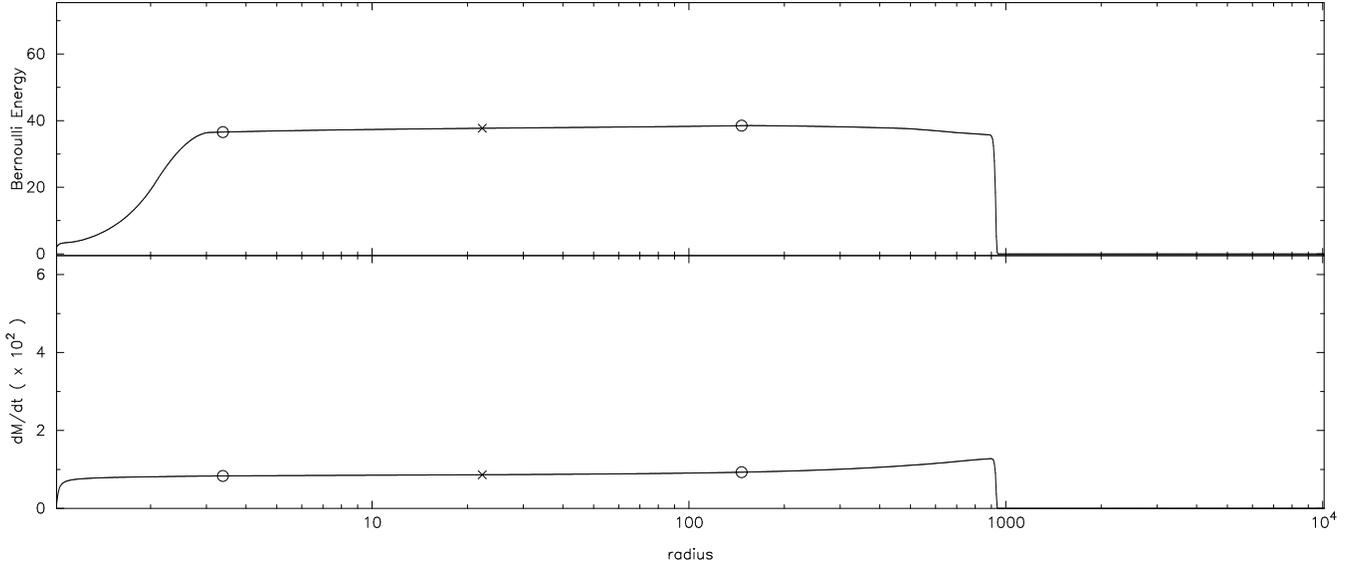}\end{turn}
\caption{The relativistic Bernoulli energy $E_{\rm
rel}=\frac{1}{2}\Gamma h/c^2 - \frac{1}{2}c^2 $(top) and mass outflow
rate $\dot{M} = 4\pi r^2\rho U^r$ (bottom) in the time-dependent
relativistic wind solution with a reasonably high heating rate are
shown as functions of radius at time $t= 1000$. In order to match this solution to a
steady outflow solution, the Bernoulli energy is assumed to be
constant over the region indicated by the two circles, and the steady
wind solution is computed using initial values at the point indicated
by a cross.}
\label{fig:relmdoten}
\end{center}
\end{figure}

 The results of a typical (n=2.0) relativistic simulation are shown in
Figure \ref{fig:relresults} at $t=1000$. Again we observe that the
wind structure reaches a quasi-steady state, with the velocity
approaching a steady value at large radii.  Note that because the steady state
density is higher than that of the surrounding medium no wide shock front is
observed.

Plotting the mass outflow rate $\dot{M} = 4\pi r^2\rho U^r$ and the
relativistic Bernoulli energy $E_{\rm rel} = \frac{1}{2}\Gamma^2
h^2/c^2 - \frac{1}{2}c^2$ (see e.g.  \citealt{st83}) as a function of
radius (Figure \ref{fig:relmdoten}), we see that indeed the structure
approaches that of a steady (relativistic) wind (that is, the energy and
$\dot{M}$ profiles are flat above the heating zone). We may thus apply a relativistic steady wind solution
with this Bernoulli energy as an initial value to determine the final
velocity and Lorentz factor as $r \rightarrow \infty$. Note that we
cannot apply a non-relativistic steady wind solution because although
the gravity is non-relativistic, the outflow velocities are not. As in
the non-relativistic case the final wind velocity is determined by the
steady Bernoulli energy, since in this case as $r \rightarrow \infty$,
$E_{\rm rel} \rightarrow \frac{1}{2}[(U^r)^2 -c^2]$.

\subsection{Steady wind solution}
\begin{figure}[t]
\begin{center}
\begin{turn}{270}\epsfig{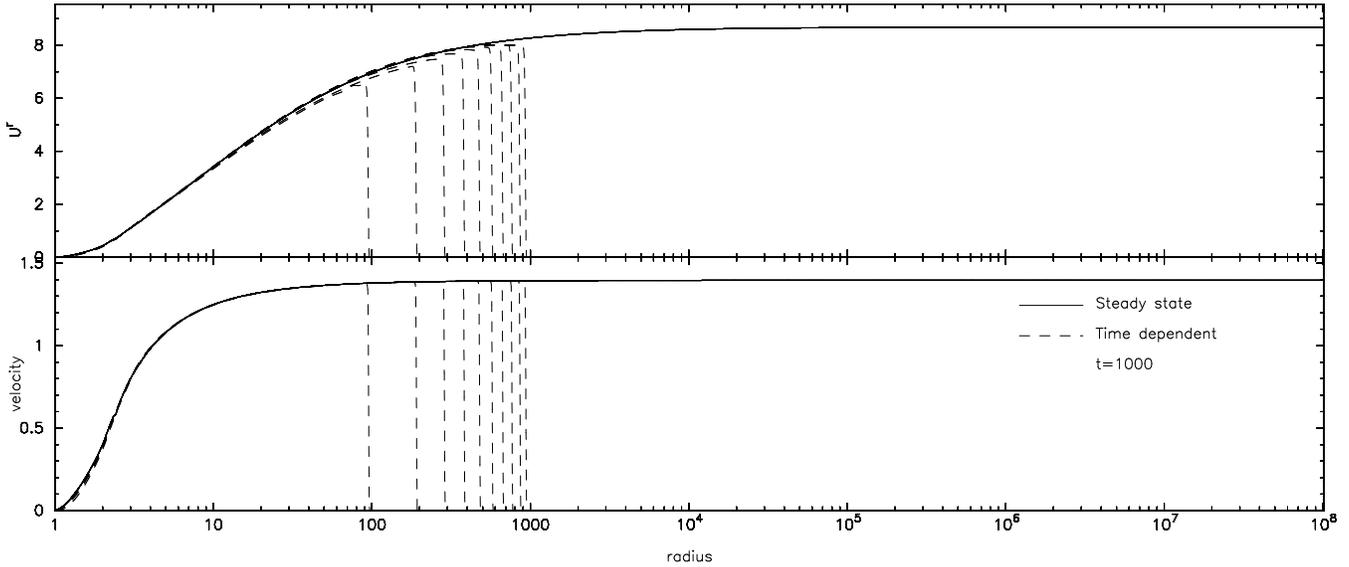}\end{turn}
\caption{The radial profiles of the steady wind r-component of four
velocity $U^r$ (top panel) and of the velocity for Eulerian observers
$\bar{v}^r$ (centre panel) are compared to the time-dependent solution
(plotted every 100 dynamical times) for a typical relativistic
calculation for the black hole (n=2.0) case. Units are such that
$c=\sqrt{2}$ on the velocity plots. Note the excellent agreement between the two
solutions.}
\label{fig:relsteady}
\end{center}
\end{figure}

 Relativistic, steady state ($\partial/\partial t = 0$) winds were
first studied by \citet{m72} and extended to include energy deposition
by \citet{f82}. The problem has recently received attention in the
context of neutrino-driven winds in gamma-ray burst models by
\citet{pfc01} and \citet{tbm01}. We proceed in a manner analogous to
that of the non-relativistic solution. Setting $\partial/\partial t
=0$ the continuity (\ref{eq:relcty1}) and momentum (\ref{eq:relmom1})
equations become
\begin{eqnarray}
\frac{1}{\rho}\pder{\rho}{r} + \frac{1}{U^r}\pder{U^r}{r} + \frac{2}{r} & = & 0 \label{eq:relstcty} \\
U^r\pder{U^r}{r} + \frac{\Gamma^2 c^2}{\rho h}\pder{P}{r} + \frac{GM}{r^2} & = & 0 \label{eq:relstmom}
\end{eqnarray}
where (\ref{eq:relstcty}) is equivalent to
\begin{equation}
r^2 \rho U^r = \mathrm{const}.
\end{equation}
 Combining (\ref{eq:relstmom}) and (\ref{eq:relstcty}) we obtain
\begin{equation}
\frac{1}{U^r}\left[(U^r)^2 - \frac{c^2\Gamma^2 c_s^2}{h\gamma}\right]\pder{U^r}{r} = -\frac{c^2\Gamma^2}{h\gamma}\frac{dc_s^2}{dr} + \frac{c^2\Gamma^2}{h\gamma}\frac{2c_s^2}{r} - \frac{GM}{r^2},
\label{eq:momcs2}
\end{equation}
where $c_s^2 = \gamma P/\rho$ and $(U^r)^2 \equiv U^rU^r$. From the first law of thermodynamics and
(\ref{eq:relstmom}) we derive the relativistic Bernoulli equation in the form
\begin{equation}
\frac{d}{dr}\left(\frac{1}{2}\frac{\Gamma^2 h^2}{c^2}\right) = \frac{h\Gamma^2}{c^2}\frac{dQ}{dr},
\label{eq:relbernoulli}
\end{equation}
such that both sides reduce to their non-relativistic expressions as
$c\to\infty$. The quantity $dQ/dr$ is the local heating gradient as in
the non-relativistic case. Expanding this equation we find
\begin{equation}
\frac{dc_s^2}{dr} = (\gamma-1)\left[\frac{dQ}{dr} -
\frac{h}{c^2\Gamma^2}\frac{d}{dr}\left(\frac{1}{2}(U^r)^2\right) -
\frac{h}{c^2\Gamma^2}\frac{GM}{r^2}\right].
\label{eq:relcssq}
\end{equation}
Combining (\ref{eq:relcssq}) and (\ref{eq:momcs2}) and manipulating terms, we obtain an equation for $(U^r)^2$,
\begin{equation}
\frac{d}{dr}(U^r)^2 = \frac{2(U^r)^2}{\left[(U^r)^2 - c^2\Gamma^2 c_s^2/h\right]}\left[\frac{c^2\Gamma^2}{h}\frac{2c_s^2}{r}
- (\gamma-1)\frac{c^2\Gamma}{h}\left(\Gamma\frac{dQ}{dr}\right) -
\frac{GM}{r^2}\right],
\label{eq:ursq}
\end{equation}
where $c_s^2$ and $h = c^2 + c_s^2/(\gamma-1)$ are given functions of
known variables by integration of the Bernoulli equation
(\ref{eq:relbernoulli}), in the form
\begin{equation}
\frac{d}{dr}\left(\Gamma h\right) = \Gamma\frac{dQ}{dr},
\end{equation}
to ensure that $h$ does not appear in the heating term on the right hand side. The integration is then
\begin{equation}
e(r) = \Gamma h = e(r_\infty) - \int^{r_\infty}_r \left\{\Gamma\frac{dQ}{dr}\right\} dr,
\end{equation}
and hence
\begin{equation}
h = \frac{e(r)}{\Gamma},\hspace{5mm} c_s^2 = (\gamma-1)(h - c^2).
\end{equation}
 The `heating gradient', $\Gamma dQ/dr$, is calculated from the time-dependent solution using
\begin{equation}
\Gamma\frac{dQ}{dr}(r) = \frac{\alpha(r)\Lambda(r)}{\bar{v}^r(r)},
\end{equation}
since 
\begin{equation}
\Lambda \equiv T\frac{ds}{d\tau} \equiv \frac{dQ}{d\tau} = U^t \left(\pder{Q}{t} + v^r\frac{dQ}{dr}\right),
\end{equation}
 where $\tau$ is the proper time and $U^t = \Gamma/\alpha^2$. The velocity
profile for an Eulerian observer is then calculated using (\ref{eq:veul})
and the final Lorentz factor $W_\infty$ using equation (\ref{eq:Lfac}). As in
the non-relativistic case we choose a starting point for the integration above
the heating shell and integrate outwards from this point using a fourth order
Runge-Kutta integrator in order to determine
the terminal Lorentz factor. The inward integration (and thus the
determination of the steady state heating gradient $\Gamma dQ/dr$) is computed
only for consistency. We integrate through the singular point in equation (\ref{eq:ursq}) by taking a
low order integration with larger steps as this point is approached.

The solution calculated using (\ref{eq:ursq}) is shown in Figure
\ref{fig:relsteady} plotted against the evolving time-dependent solution. The
profiles are in excellent agreement, verifying the accuracy of the relativistic
calculation and showing that the wind may indeed be described by the steady state
solution.

\subsection{Terminal wind velocities and Lorentz factors as a function of
heating rate}
\begin{figure}[t]
\begin{center}
\begin{turn}{270}\epsfig{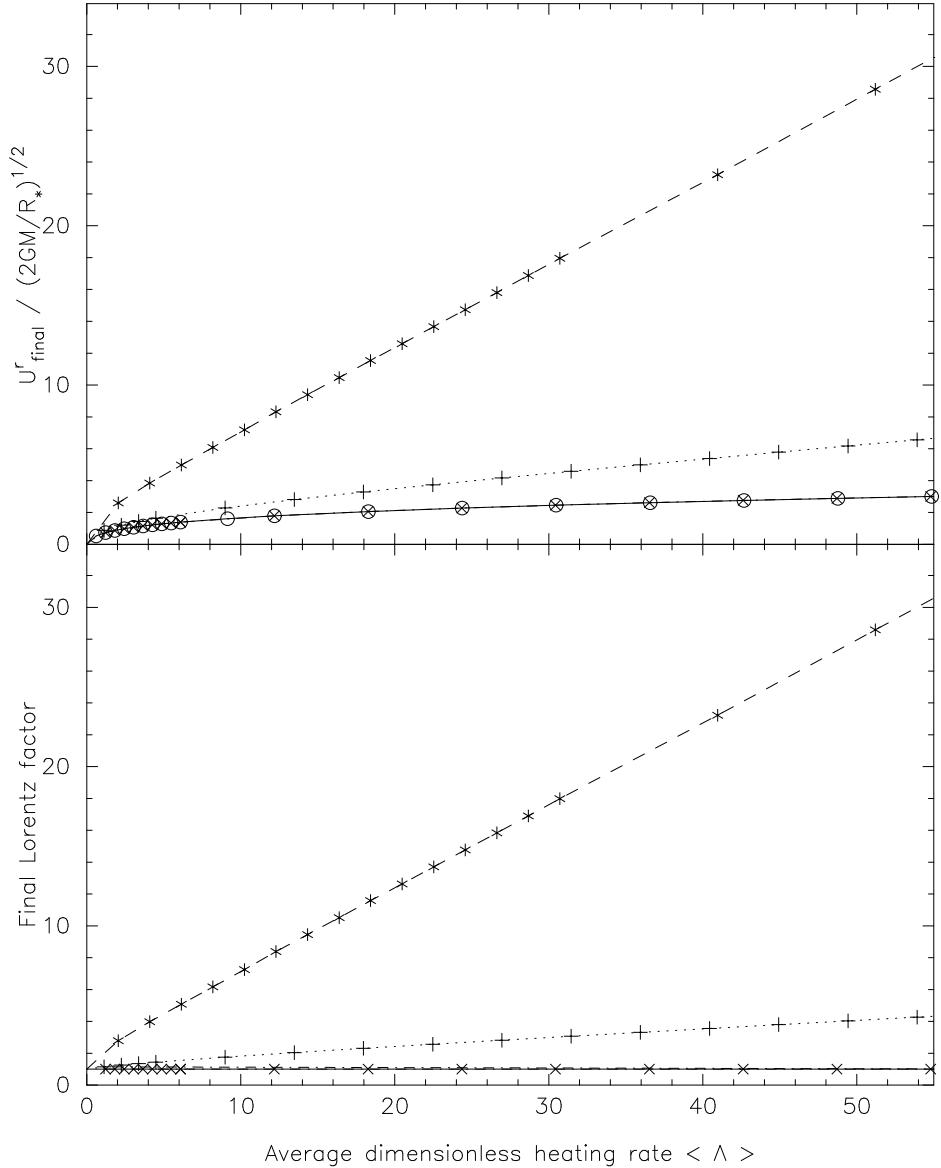}\end{turn}
\caption{The terminal r-component of four velocity $U^r$ (top panel)
and Lorentz factor (bottom panel) of the wind in the non-relativistic
($\circ$, solid), white dwarf ($\times$, dot-dashed), neutron star ($+$,
dotted) and black hole
($*$, dashed) cases, is plotted as a function of the dimensionless heating rate
defined in $\S\ref{sec:nrrates}$. The top panel may be compared with
Figure~$\ref{fig:nrrates}$
in the non-relativistic case.}
\label{fig:relrates}
\end{center}
\end{figure}

In order to compare the relativistic results to those in the Newtonian regime,
we define the local canonical heating rate in a similar manner to
the non-relativistic case, that is
\begin{equation}
\Lambda_c(r) = \frac{\Delta E}{\Delta t},
\end{equation}
for some relevant energy $\Delta E$ and some relevant timescale
$\Delta t$. As in Section~\ref{sec:nrrates} we take the canonical energy per
unit mass, $\Delta E$, to be the energy released locally by bringing
to rest a particle of unit mass which is orbiting in a circular orbit
at radius $r$. For a particle orbiting in the Schwarzschild metric
this is the difference, $\Delta E$, between the energy constants
(defined by the timelike Killing vector) of a circular geodesic at
radius $r$, and a radial geodesic with zero velocity at radius
$r$. This implies (see, for example, \citealt{s90}, Chapter 11)
\begin{equation}
\Delta E/ c^2  =  \frac{1 - 2GM/rc^2}{[1 - 3GM/rc^2]^{1/2}} - [1 -
2GM/rc^2]^{1/2}.
\end{equation}
In the Newtonian limit, this reduces to the expected value $\Delta E =
\frac{1}{2} v_{\phi}^2 = GM/2r$. We again take the canonical timescale
on which the energy is released to be the orbital timescale at radius
$r$ as measured by a local stationary observer. For a circular
geodesic in the Schwarzschild metric, the azimuthal velocity is given
in terms of coordinate time, $t$, by
\begin{equation}
d \phi / d t = \Omega = (GM/r^3)^{1/2}.
\end{equation}
This is the same expression as for the angular velocity of an orbiting
particle in the Newtonian limit.  But in terms of the proper time,
$\tau$, of a local stationary observer we have, from the metric, 
\begin{equation}
d \tau / d t = (1 - 2GM/rc^2)^{1/2},
\end{equation}
and thus $d \phi/ d \tau = \Omega_o$, where
\begin{equation}
\Omega_o^2 = \frac{GM}{r^3} \left[ 1 - \frac{2GM}{rc^2}\right]^{-1}.
\end{equation}
Using this, the local canonical heating rate is therefore given by
\begin{equation}
\Lambda_c = \Delta E  \times \Omega_o.
\end{equation}
In the Newtonian limit, $r \gg 2GM/c^2$, this becomes as expected
$\Lambda_c \simeq (GM)^{3/2}/2 r^{5/2}$. As in the non-relativistic
case we use the canonical heating rate derived above to define a
dimensionless heating rate $\langle \Lambda \rangle$ as an appropriate
volume average using equation~(\ref{eq:lambdaav}).

The final Lorentz factor of the wind plotted as a function of this
dimensionless heating rate is given in the bottom panel of Figure
\ref{fig:relrates} in the highly relativistic (black hole), moderately
relativistic (neutron star, equivalent to a broader heating shell
further away from a black hole) and non-relativistic (white dwarf)
cases. 

 We would also like to make a meaningful comparison of the final wind
velocities in units of the escape velocity from the star. Note that we
cannot simply compare the scaled velocities since we are in effect
introducing a `speed limit' in the relativistic solution such that the
(scaled) relativistic velocity will always be slower than in the
equivalent non-relativistic solution. However we can compare the
velocity for observers along the worldline of a particle in the wind
(ie. observers with proper time interval $d\tau$), $U^r = dr/d\tau$
(that is, the $r$ component of the four velocity, which in special
relativity is given by $U^r = \gamma v^r$, where $\gamma$ is the
Lorentz factor). Scaling this in units of the (Newtonian) escape
velocity from the central object $(2GM/R_*)^{1/2}$ we can make a
useful comparison with the non-relativistic results. This velocity is
plotted in the top panel of Figure \ref{fig:relrates} against the
dimensionless heating rate.

\section{Discussion and Conclusions}
\label{sec:discussion}

We have considered the input of energy at the base of an initially
hydrostatic atmosphere as a simple model for the acceleration of an
outflow or jet. The problem is inherently a time-dependent one,
because the flow velocity at the base of the atmosphere is
zero. Sufficiently large energy input rates give rise to supersonic
outflows. We are, of course, unable to compute the outflow for an
infinite time, and thus cannot directly measure the terminal outflow
speed. However, we make use of the fact that if the mass in the
atmosphere is sufficiently large compared to the mass outflow rate,
then at large radii the outflow approximates to a steady state (with
constant mass flux). We thus match our time-dependent solutions onto
steady state outflow solutions at large radii and thus determine the
terminal velocities for the outflows. We then compute how the terminal
velocity of the outflow varies as a function of (dimensionless)
heating rate in both Newtonian and relativistic gravitational
potentials. The results are shown in Figures~5 and~\ref{fig:relrates}.

In the top half of Figure~\ref{fig:relrates} we note that
dimensionless energy (or momentum) imparted to the outflow at a given
value of the dimensionless heating rate is larger in the relativistic
regime. This comes about simply because a particular element of gas
cannot escape from the zone in which the heating is occurring with a
velocity which exceeds the speed of light. Thus a gas element which is
accelerated to relativistic energies in the heating zone spends longer
in the heating zone than one which is not. Indeed from Figure~10, we
see that, once the outflows become relativistic, the energy per unit
mass in the outflow (proportional to $\gamma$) is proportional to the
dimensionless heating rate $\langle \Lambda \rangle$. This comes about
because each fluid element spends the same time in the heating zone,
because it travels through the zone at a velocity $\sim c$. 

 From Figure~5 we see that a dimensionless heating rate of $\langle
\Lambda \rangle \simeq 17$ gives rise to a terminal outflow velocity
of $v_{\rm jet} \simeq 2 v_{\rm esc}$ in a Newtonian potential. From
Figure~10, we see that for the same heating rate, the `neutron star'
wind, for which the heating rate peaks at about 5.2 $R_{\rm Sch}$ becomes
mildly relativistic ($\gamma_{\rm jet} \sim 2$), whereas the `black
hole' wind, for which the heating rate peaks at about 2.1 $R_{\rm
Sch}$, leads to an outflow with $\gamma_{\rm jet} \simeq 11$. Similarly
a dimensionless heating rate of $\langle \Lambda \rangle \simeq 55$
gives rise to a terminal velocity of $v_{\rm jet} \simeq 3 v_{\rm
esc}$ in the Newtonian case, to an outflow with $\gamma_{\rm jet} \sim
4$ in the mildly relativistic case, and to an outflow with $\gamma_{\rm
jet} \simeq 31$ in the strongly relativistic case. We have noted above
that although the exact numerical values here do depend slightly on
the exact definition of the dimensionless heating rate, the basic
results remain unchanged.  For example, using the Newtonian
dimensionless heating rate (\S\ref{sec:nrrates}) in the strongly
relativistic case gives a Lorentz factor of $\gamma_{\rm jet} \simeq
5$ for the rate which corresponds to $v_{\rm jet} \simeq 2 v_{\rm
esc}$ in the non-relativistic case.

We caution that the above analysis does not demonstrate that the model
we use provides an adequate description of the physics involved in the
acceleration process (for example, the jet energy might be initially
mainly in electromagnetic form -- Poynting flux -- and only later
converted to kinetic energy of baryons) or that there is no intrinsic
difference between the jet outflows caused by the relativistic nature
of the AGN jets (for example, the AGN jets might be mainly in the form
of a pair plasma and therefore lighter).  And we note that it is
evident that more detailed physical models need to be developed before
further conclusions can be drawn. Nevertheless, we suggest that the
generic nature of our analysis might give some insight into the
physical processes involved in the acceleration of jets.

Thus we conclude, on the basis of the rather simplified physical model
we have employed in our analysis, that it is not unreasonable to argue
that the jets in AGN are simply scaled up, relativistic versions of
the jets in YSOs, and that the intrinsic jet acceleration mechanism is
indeed the same in both the AGN and the YSO contexts. In making this
analogy, again on the basis of our simplified model for the
acceleration process, we find two further physical conditions must
hold. First, we find that the energy input process, which leads to the
acceleration of the outflow, takes place deep in the (relativistic)
gravitational well for the AGN case ($\sim 2 R_{Sch}$). What this means physically is
that in this way it is possible to make use of the limiting velocity,
$c$, to ensure that relativistic fluid elements remain relatively
longer in the energy input zone compared to their non-relativistic
counterparts. Second, we find that the required dimensionless heating
rate is much larger than unity. The physical implication of this is
that the available energy released in the accretion process must be
imparted to a small fraction of the available accreting material.

\section*{Acknowledgements}
DJP acknowledges the support of the Association of Commonwealth
Universities and the Cambridge Commonwealth Trust.  He is supported by
a Commonwealth Scholarship and Fellowship Plan. JEP acknowledges
useful discussions with Dr M. Livio. DJP and JEP also acknowledge Dr R. Carswell
for useful discussion.

\appendix
\section{Discretization scheme for non-relativistic equations}
\label{sec:appendixA}
\begin{figure}[h]
\begin{center}
\begin{turn}{0}\epsfig{file=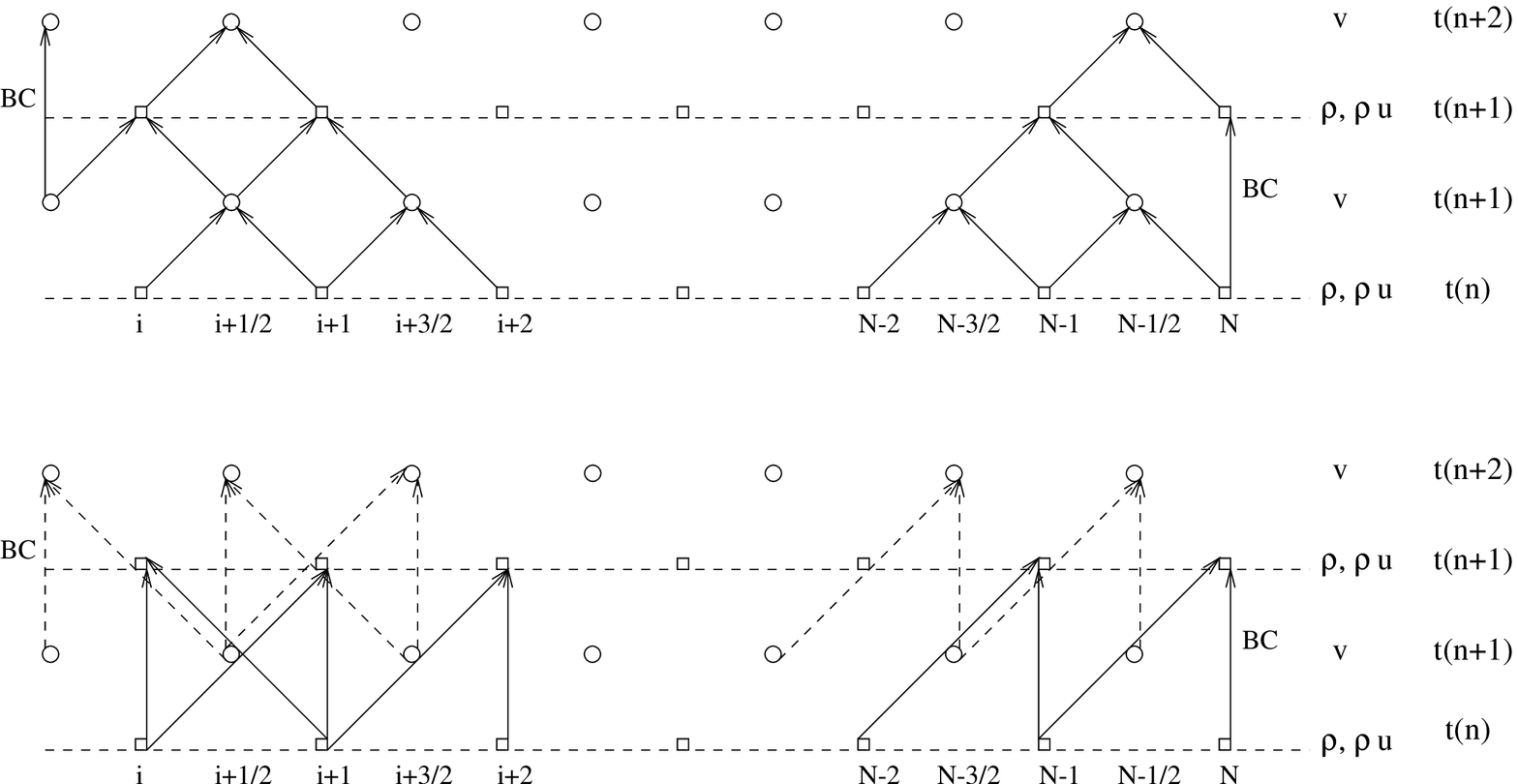,width=1.0\textwidth}\end{turn}
\caption{Schematic diagram of numerical method: density and internal energy are
defined on the integer points while velocity is calculated on the half points.
The solution requires one inner boundary condition on $v$ and two outer
boundary conditions for $\rho$ and $\rho u$. Updated velocities ($v^{n+1}$) are used to
calculate $\rho^{n+1}$ and $\rho u^{n+1}$. The scheme allows centred
differencing on terms involving staggered quantities (top panel) while upwind
differencing is used on the advective terms (bottom panel).}
\label{fig:scheme}
\end{center}
\end{figure}

The discretization scheme for the non-relativistic fluid equations is
summarised in Figure \ref{fig:scheme}. Fluxes are calculated on the half grid
points while the other terms are calculated on the integer points. We
solve (\ref{eq:nrcty})-(\ref{eq:nreos}) in the following manner: The numerical equations are solved first
for velocity on the half grid points (dropping the superscript $r$ for
convenience),
\begin{eqnarray}
v^{n+1}_{i+1/2} & = & v^{n+1}_{i+1/2} - \Delta t\left[
v^n_{i+1/2}\left(\frac{v^n_{i+3/2}-v^n_{i+1/2}}{r_{i+3/2} - r_{i+1/2}}\right) -
\frac{1}{\rho^n_{i+1/2}}\left(\frac{P^n_{i+1}-P^n_i}{r_{i+1} - r_i}\right) -
\frac{1}{r^2_{i+1/2}}\right] \hspace{5mm}(v<0)\nonumber \\ 
& = &  v^{n+1}_{i+1/2} - \Delta t\left[ v^n_{i+1/2}\left(\frac{v^n_{i+1/2}-v^n_{i-1/2}}{r_{i+3/2} -
r_{i+1/2}}\right) - \frac{1}{\rho^n_{i+1/2}}\left(\frac{P^n_{i+1}-P^n_i}{r_{i+1} - r_i}\right) -
\frac{1}{r^2_{i+1/2}} \right] \hspace{5mm}(v>0)
\end{eqnarray}
where the superscript $n$ refers to the $n$th timestep and the subscript $i$
refers to $i$th grid point ($v_{i+1/2}$,$\rho_{i+1/2}$ thus being points on the staggered
velocity grid). The quantity $\rho_{i+1/2}$ is calculated using linear
interpolation between the grid points, ie. $\rho_{i+1/2} =
\frac{1}{2}(\rho_i+\rho_{i+1})$. We then solve for the density and internal
energy on the integer grid points using the updated velocity,
\begin{eqnarray}
\rho^{n+1}_i & = & \rho^n_i - \Delta t\left[ v^{n+1}_i\left(\frac{\rho^n_{i+1}-\rho^n_i}{r_{i+1}-r_i}\right) -
\frac{\rho_i^n}{r_i^2}\left(\frac{r^2_{i+1/2}v^{n+1}_{i+1/2} -
r^2_{i-1/2}v^{n+1}_{i-1/2}}{r_{i+1/2}-r_{i-1/2}}\right)\right] \hspace{5mm}(v<0)\nonumber \\
 & = & \rho^n_i - \Delta t\left[ v^{n+1}_i\left(\frac{\rho^n_i-\rho^n_{i-1}}{r_i-r_{i-1}}\right) -
\frac{\rho_i^n}{r_i^2}\left(\frac{r^2_{i+1/2}v^{n+1}_{i+1/2} -
r^2_{i-1/2}v^{n+1}_{i-1/2}}{r_{i+1/2}-r_{i-1/2}}\right)\right] \hspace{5mm}(v>0)
\end{eqnarray}
and similarly,
\begin{eqnarray}
\rho u^{n+1}_i & = & \rho u^n_i - \Delta t\left[v^{n+1}_i\left(\frac{\rho u^n_{i+1}-\rho
u^n_i}{r_{i+1}-r_i}\right) - \left[\frac{P_i^n + \rho u^n_i}{r_i^2}\right]\left(\frac{r^2_{i+1/2}v^{n+1}_{i+1/2} -
r^2_{i-1/2}v^{n+1}_{i-1/2}}{r_{i+1/2}-r_{i-1/2}}\right) + \rho^n_i \Lambda_i
\right] \hspace{5mm}(v<0) \nonumber\\
 & = & \rho u^n_i - \Delta t\left[v^{n+1}_i\left(\frac{\rho u^n_i-\rho
u^n_{i-1}}{r_i-r_{i-1}}\right) - \left[\frac{P_i^n + \rho u^n_i}{r_i^2}\right]\left(\frac{r^2_{i+1/2}v^{n+1}_{i+1/2} -
r^2_{i-1/2}v^{n+1}_{i-1/2}}{r_{i+1/2}-r_{i-1/2}}\right) + \rho^n_i \Lambda_i \right] \hspace{5mm}(v>0)
\end{eqnarray}
where $\Delta t = t^{n+1} - t^{n}$ and the timestep is regulated according to the Courant condition
\begin{equation}
\Delta t < \frac{\mathrm{min}(\Delta r)}{\mathrm{max}(\left|{v}\right|) + \mathrm{max}(c_s)}
\end{equation}
where $c_s$ is the adiabatic sound speed in the gas given by $c_s^2 = \gamma P/\rho$. We typically set
$\Delta t$ to half of this value.

\label{lastpage}
\end{document}